\documentclass[%
superscriptaddress,
bibnotes,
 amsmath,amssymb,
 aps,
 prl, 
eprint,
floatfix,
reprint
]{revtex4-2}
\usepackage{graphicx,xcolor}
\definecolor{darkblue}{RGB}{0,0,150}
\definecolor{nightblue}{RGB}{0,0,100}

\usepackage{mathrsfs,dsfont,mathtools}

\newcommand{\refsub}[2]{\hyperref[#1]{\ref*{#1}#2}}

\usepackage{dcolumn}
\usepackage{bm}
\usepackage[
colorlinks,
citecolor=darkblue,
linkcolor=darkblue,
urlcolor=nightblue]{hyperref}

\usepackage[english]{babel}
\usepackage[babel,kerning=true,spacing=true]{microtype}

\newcommand{\bk}{\mathbf{k}}

\definecolor{DarkRed}{RGB}{100,0,0}
\definecolor{LightGreen}{RGB}{000,50,0}

\begin{document}

\title{
Mixed axial-gravitational anomaly from emergent curved spacetime\protect\\ in nonlinear charge transport}

\author{Tobias Holder}
\email{tobias.holder@weizmann.ac.il}
\author{Daniel Kaplan} 
\affiliation{Department of Condensed Matter Physics,
Weizmann Institute of Science, Rehovot, Israel}
\author{Roni Ilan}
\affiliation{Raymond and Beverly Sackler School of Physics and Astronomy, Tel Aviv University, Tel Aviv, Israel}
\author{Binghai Yan}
\email{binghai.yan@weizmann.ac.il}
\affiliation{Department of Condensed Matter Physics,
Weizmann Institute of Science, Rehovot, Israel}
\date{\today}

\begin{abstract}
In 3+1 dimensional spacetime, two vector gauge anomalies are known: The chiral anomaly and the mixed axial-gravitational anomaly. While the former is well documented and tied to the presence of a magnetic field, the latter instead requires a nonzero spacetime curvature, which has made it rather difficult to study.
In this work, we show that a quantum anomaly arises in the second-order electrical response for zero magnetic field, which creates a dc-current that can be either longitudinal or transverse to electric field. 
Consequently, the continuity equation for the chiral current is not conserved at order $\tau^{-1}$, where $\tau$ is the quasiparticle relaxation time.
We can identify the anomaly as a mixed axial-gravitational one, and predict a material in which the anomaly-induced current can be isolated in a purely electrical measurement. 
Our findings indicate that charge transport generically derives from quasiparticle motion in an emergent curved spacetime, with potentially far-reaching consequences for all types of response functions. 
\end{abstract}

\maketitle

\begin{figure}[t]
    \centering
    \includegraphics[width=\columnwidth]{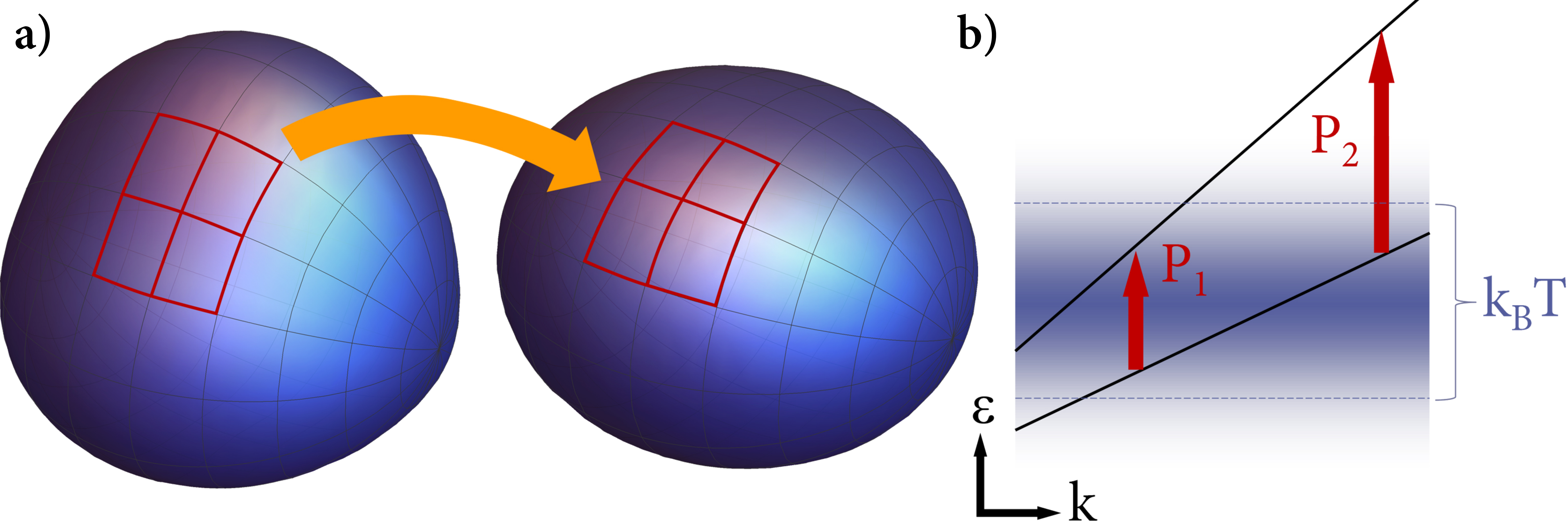}
    \caption{
    (a) Origin of the anomaly according to semiclassics, where all processes are manifest as Fermi surface effects:
    The applied field renormalizes the Fermi surface,
    thereby changing the local low-energy manifold of a given state, which redistributes carriers between different parts of the Fermi surface, or even between different Fermi surfaces, thus violating charge conservation within a given patch of momentum space, although the total charge is conserved. 
    (b) 
    Origin of the mixed axial-gravitational anomaly according to perturbation theory:
    Carriers are moved from the Fermi surface into remote bands, thus violating charge conservation at the Fermi level. Such is possible via thermal and non-thermal transitions (blue: thermal window around the chemical potential). 
    At finite temperature, real transitions of type $P_1$ can take place which move carriers from or to the Fermi level. The same process can take place virtually ($P_2$) at second order in the electric field, without the need for a real occupation in the excited band. The anomaly therefore persists at temperature $T=0$. 
    }
    \label{fig:fig1}
\end{figure}

When cold electrons move in a perfect, static and stable lattice, according to textbook wisdom, they move like almost free quasiparticles in flat space. 
Disregarding 'dirt'-effects, one might even be tempted to ask, given an excited but  low-energy state in vicinity of the Fermi surface, how much does the lattice really matter? Since several years, the impactful insight has been that yes, the lattice matters on a fundamental level due to nontrivial topology present in the band structure~\cite{Xiao2010,Hasan2010,Qi2011}.
However, this insight sidelines the more subtle question how much the lattice matters \emph{locally} and \emph{dynamically}. Why do electronic quasiparticles actually move in flat space, given that they navigate through a dense forest of lattice sites?
The answer is simply that quasiparticles do \emph{not} move in a flat space, not even in a perfect and static lattice at zero temperature. 
To substantiate this statement, here we make use of a quantum anomaly in a purely electric, but nonlinear conductivity without thermal gradients or magnetic field.

Quantum anomalies appear when a classically conserved quantity is not conserved on the quantum level, but broken by quantum fluctuations~\cite{Bertlmann2000}. 
They play a central role in the theoretical and experimental understanding of quantum matter, 
because they imply the existence of non-conserved processes which would be forbidden classically. 
While anomalies were first discussed in the standard model of elementary particle physics~\cite{Adler1969,Bell1969}, condensed matter systems like Weyl semimetals can provide accessible experimental platforms to study their effects~\cite{Nielsen1983,Son2013}.
Weyl semimetals are three-dimensional compounds which feature pairs of massless chiral fermions~\cite{Yan2017,Armitage2018}. In the presence of a magnetic field Weyl fermions exhibit a chiral anomaly, which means that the quasiparticle number of each chirality is not conserved. 
By now, the chiral anomaly in Weyl semimetals is well understood theoretically~\cite{Son2012,Son2013,Landsteiner2014,Kharzeev2014} and confirmed experimentally by detecting negative magnetoresistance (see Refs.~\cite{Yan2017,Armitage2018,Lv2021} for a review).

Massless chiral fermions additionally exhibit a mixed axial-gravitational anomaly (AGA), which in curved spacetime leads to a non-conservation of the chiral charge and of the energy-momentum tensor~\cite{Delbourgo1972,Eguchi1976,AlvarezGaume1984}.
Due to its gravitational origin, the AGA is much harder to observe. 
One approach to probe the AGA is offered in a hydrodynamic picture, where a nonuniform velocity profile can be interchanged for a gravitomagnetic potential, i.e. for an emergent curved spacetime. This is known as the chiral vortical effect~\cite{Landsteiner2011}. 
However, in recent years, it became clear that even in the standard Kubo formalism (in flat space) the magnetic field dependence of thermoelectric transport coefficients is modified by the AGA~\cite{Landsteiner2011,Lucas2016a,Lundgren2014,Kim2014,Sharma2016,Spivak2016,Gooth2017}.
This finding is reminiscent of Luttinger's suggestion to use a gravitational potential to calculate thermal transport coefficients~\cite{Luttinger1964}.
However, while thermal effects obviously originate from a modification of the occupation numbers, 
the conjectured connection of occupation changes with gravitational effects has remained mostly anecdotal~\cite{Tatara2015,Xiao2020,Park2021a}.

We point out that AGA must also be visible upon manually changing occupation numbers, even if the perturbed system is not given by a thermal ensemble, if there exists an equivalence between changes in occupation and the emergence of a nontrivial spacetime metric (i.e. emergence of non-Euclidian spacetime).
In this work, we show how to draw such an inference using the example of nonlinear charge transport, which exhibits an unambiguous AGA with respect to conservation of chiral charge. 
To this end, we first point out that the second-order electrical conductivity couples quadratically to the electric field, just like it is the case for Joule heating.
In both cases, carriers undergo vertical (i.e. finite energy) transitions, i.e. changes to occupation numbers. These transitions are of thermal nature in case of Joule heating while they are virtual for the second-order conductivity (Fig.~\ref{fig:fig1}).
Consequently, the second-order conductivity should couple to an emergent nontrivial spacetime metric, if it exists at all.
The appearance of the AGA might alternatively be viewed as resulting from a renormalization of the shape of the Fermi surface which locally in momentum space vacates or adds carriers (cf. Fig.~\ref{fig:fig1}).
We stress that the AGA in the second order conductivity does not imply the appearance of concomitant thermoelectric or thermal currents, and emphasize that the second order conductivity is a purely electrical conductivity.

A much more detailed justification why and how the nonlinear conductivity implies a nontrivial effective metric is given elsewhere~\cite{Holder2021}. 
Here, we focus on the AGA caused by motion in a curved spacetime as it is encoded in the second order dc-conductivity $\sigma^{(2)}$, corresponding to a static nonlinear current $j_c =\sigma^{(2)}_{ab;c}E_aE_b$ in response to an applied electric field $\bm{E}$.
We reveal that $\sigma^{(2)}$ manifests the AGA by using standard perturbation theory.
We further derive the corresponding anomalous continuity equation for the chiral current in a semiclassical picture.
As material candidates to observe the anomaly, we suggest a certain class of antiferromagnetic Dirac semimetals which break both spatial inversion ($P$) and time-reversal symmetry ($T$), while preserving $\mathcal{PT}$. In these materials,  conventional intrinsic and extrinsic Hall effects vanish due to $\mathcal{PT}$, making it easier to isolate the anomaly effect. 

\emph{Triangle diagram.---}
\begin{figure}[t]
    \centering
    \includegraphics[width=.8\columnwidth]{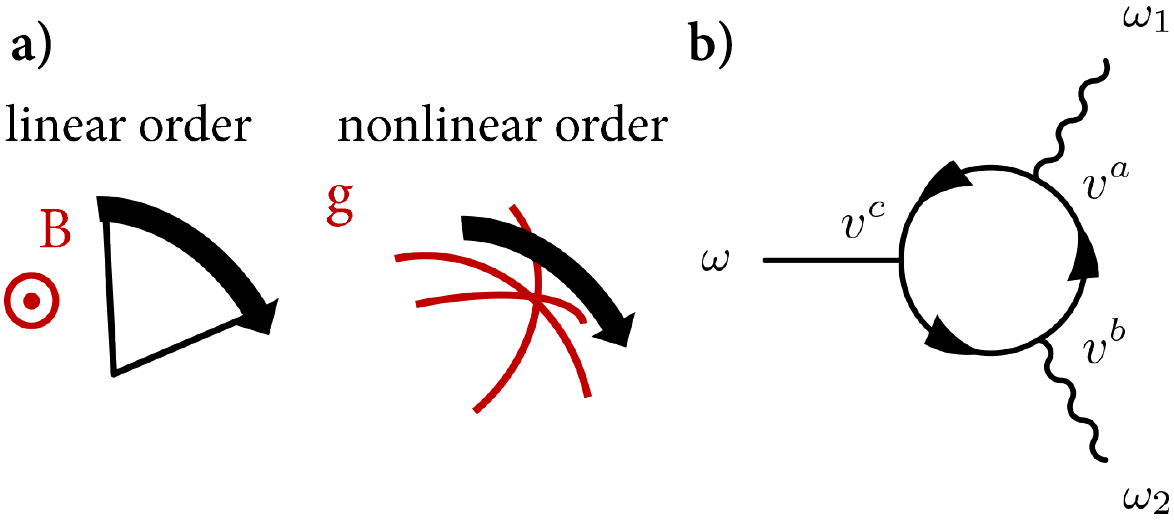}
    \caption{
    (a) Semiclassical view of the anomalous motion: At linear order, the quasiparticle acquires an anomalous velocity, which is transverse to the dispersive velocity. At second order, one obtains an anomalous acceleration, which is a result of the nontrivial manifold in which the quasiparticle moves. (b) Triangle diagram which produces the anomaly at second order in the electric field. The current is probed at frequency $\omega$, whereas the electric fields have frequencies $\omega_1$ and $\omega_2$. 
    }
    \label{fig:diagram}
\end{figure}
The study of second order conductivity in terms of canonical perturbation theory has a long history~\cite{Belinicher1980,vonBaltz1981}.
For metallic systems three contributions to $\sigma^{(2)}=\sigma^{dr}+\sigma^{bc}+\sigma^{gr}$ are known~\cite{Gao2014,Sodemann2015,Gao2019,Watanabe2020}.
$\sigma^{dr}$ is a purely dispersive term, while $\sigma^{bc}$ is produced by the so-called Berry curvature dipole~\cite{Sodemann2015,Zhang2018a}. 
In the following we will argue that the third term, $\sigma^{gr}$ originates from the AGA.

The diagrammatic description of second order optical response has been developed only much later~\cite{Parker2019,Holder2020}. 
The latter formulation allows to identify the underlying physical processes that create a nonlinear current, and also clarifies the role of finite relaxation rates.
It turns out that second order optical response is encoded by the (anomalous) acceleration of the quasiparticle while it traverses the lattice~\cite{Holder2020}. Accelerations (but not velocities) are sensitive to a motion in non-Euclidian space (cf. Fig.~\refsub{fig:diagram}{a}), which already hints that gravitational effects could become important.
Quantum anomalies usually originate from triangle diagrams, and 
indeed, in the perturbation theory for second order response, triangle diagrams of the type shown in Fig.~\refsub{fig:diagram}{b} appear entering in $\sigma^{bc}$ and $\sigma^{gr}$.
The static limit is obtained from the optical response by replacing $\omega\to i/\tau$, where $\tau$ s a finite lifetime in relaxation time approximation~\cite{Passos2018,Holder2020}.

In the following, we consider a non-interacting fermionic system with band dispersion $\hbar\varepsilon_n(\bm{k})$
with band index $n$ and the cell-periodic part of the Bloch wavefunctions $|n(\bm{k})\rangle$, where the lattice momentum $\bm{k}$ takes values inside the first Brillouin zone. One can then define the complex (Berry) connection $[r_{a}(\bm{k})]_{mn}=\langle m(\bm{k})|\partial_{k_a}|n(\bm{k})\rangle$~\footnote{We note that defining the band-diagonal elements of the Berry connection require additional care. This is however immaterial for the present discussion which only involves off-diagonal elements.} and the Berry curvature 
$[\Omega_{ab}(\bm{k})]_{nn}=i\sum_{m\neq n}([r_a(\bm{k})]_{nm}[r_b(\bm{k})]_{mn}-[r_b(\bm{k})]_{nm}[r_a(\bm{k})]_{mn})$. The Fermi-Dirac function is $f_n(\varepsilon(\bm{k}))$. When possible, we will henceforth suppress the momentum and band indices.

Using perturbation theory, we obtain the three nonlinear conductivities~\cite{Kaplan2021a}
\begin{align}
	\sigma^{dr}_{ab;c}&=
	\frac{2e^3}{\hbar^2}\tau^2
	\int_\bk \sum_n f (\partial_{k_a} \partial_{k_b} \partial_{k_c} \varepsilon)
	\label{eq:tau2}\\
	\sigma^{bc}_{ab;c}&=
	\frac{2e^3}{\hbar^2}\tau^1
	\int_\bk \sum_n f \left(\partial_{k_a} \Omega_{bc} + \partial_{k_b} \Omega_{ac}\right)
	\label{eq:tau1}\\
	\sigma^{gr}_{ab;c}&=
	\frac{2e^3}{\hbar^2}\tau^0
	\int_\bk \sum_n f 
	(\partial_{k_c} G_{ab}).
	\label{eq:tau0}
\end{align}
Here, we introduced the deformation density of states (DDS), defined as
\begin{align}\label{eq:defDDS}
	[&G_{ab}(\bm{k})]_{nn}
	\notag\\
	&=\sum_{m\neq n}
	\frac{[r_a(\bm{k})]_{nm}[r_b(\bm{k})]_{mn}+[r_b(\bm{k})]_{nm}[r_a(\bm{k})]_{mn}}{\varepsilon_m(\bm{k})-\varepsilon_n(\bm{k})}.
\end{align}
The DDS resembles a momentum-resolved Landau-Zener formula~\cite{Wittig2005,Weinberg2017}. 
Very importantly, Eqs.~(\ref{eq:tau2}-\ref{eq:tau0}) identically recover the semiclassical expressions derived in the literature. 
However, only the diagrammatic approach gives insight into the permutation symmetry of the spatial indices $(a,b,c)$. Namely, in the dc-limit the quantum effective action and thus gauge-invariant perturbation theory must be invariant under permutations of $(a,b,c)$~\cite{Jensen2013,Landsteiner2016}. 
Indeed, from Eqs.~(\ref{eq:tau2}-\ref{eq:tau0}) we conclude that this is the case for $\sigma^{dr}$ and $\sigma^{bc}$ under cyclic permutations, and also for anticyclic ones after recognizing that the operation $a\leftrightarrow b$ implies $\omega\to-\omega$ (i.e. $\tau\to-\tau$). 

However, $\sigma^{gr}$ is clearly not invariant under permutations involving the index $c$ of the current.
Such a violation of permutation symmetry means that the current due to $\sigma^{gr}$ cannot be expressed as a functional variation of an effective action.
This is a sufficient criterion~\cite{Landsteiner2016} to identify $\sigma^{gr}$ as a quantum anomaly~\footnote{Because the perturbation theory respects gauge-invariance, it reproduces the covariant form of anomalies. In the consistent formulation the anomaly would retain permutation symmetry.}.
Next we show how to generate the same phenomenology in a semiclassical approach.

\emph{Continuity equation.---}
In semiclassics, interband transitions do not appear explicitly. Instead, corresponding Fermi surface quantities are introduced which encapsulate their effect. 
The paradigmatic example for this is the Berry curvature $\Omega$, a Fermi surface quantity that is however defined using information from the entire band structure. 
For orientation, first we show how the presence of $\Omega_{ab}$ introduces the chiral anomaly and how this affects the continuity equation~\cite{Son2013}.
The starting point is the kinetic equation for the distribution function $f^\pm(\bm{p},\bm{r},t)$ for each chirality in relaxation time approximation for the intervalley relaxation rate $1/\tau'$,
\begin{align}\label{eq:boltzmann}
	&(1+\tfrac{e}{c\hbar^2}\bm{B}\cdot \bm{\Omega})\partial_t f^\pm
	+(\bm{v}+\tfrac{e}{\hbar^2}\bm{E}\times \bm{\Omega}+\tfrac{e}{c\hbar^2}(\bm{\Omega}\cdot\bm{v})\bm{B})
	\partial_{\bm{r}}f^\pm
	\notag\\
	&\quad+
	(e\bm{E}+\tfrac{e}{c}\bm{v}\times\bm{B}+\tfrac{e^2}{c \hbar^2}(\bm{E}\cdot\bm{B})\bm{\Omega})\partial_{\bm{p}}f^\pm
	\notag\\
	&=-(1+\tfrac{e}{c \hbar^2}\bm{B}\cdot \bm{\Omega})\tfrac{f^\pm-f^\pm_0}{\tau'}
\end{align}
For a given chirality, integrating this over momenta yields the continuity equation, usually stated as
\begin{align}\label{eq:continuity}
	\partial_t n^\pm+\bm{\nabla}\cdot \bm{j}^\pm
	\pm\tfrac{e^2}{4\pi^2\hbar^2 c}\bm{E}\cdot\bm{B}&=
	-\frac{\delta n^\pm}{\tau'}.
\end{align}
The distribution function that solves Eq.~\eqref{eq:boltzmann}
is 
\begin{align}
f^\pm&=f^\pm_0+\frac{\tau'}{\hbar}\frac{e\bm{E}+\tfrac{e^2}{c\hbar^2}(\bm{E}\cdot\bm{B})\bm{\Omega}}{1+\tfrac{e}{c\hbar^2}\bm{B}\cdot \bm{\Omega}}\cdot\partial_{\bm{p}}f^\pm
\end{align}
This distribution function also fulfills the continuity equation, Eq.~\eqref{eq:continuity}, namely $\partial_t n^\pm=0$, $\bm{\nabla}\cdot \bm{j}^\pm=0$ and $\delta n^\pm=\int_{\bm{p}}(f^\pm-f^\pm_0)=\pm\tfrac{\tau' e^2}{4\pi^2\hbar^2 c}\bm{E}\cdot\bm{B}$. 
However, due to the anomaly, both sides of Eq.~\eqref{eq:continuity} are actually non-zero, which is a testament to the fact that in the steady state, chiral charge is continuously leaking between both chiralities.

Equipped with this intuition, we now inspect the continuity equation at second order in the electric field. 
Compared to the case of the chiral anomaly, in the following discussion we will not assume a Weyl-type dispersion with regions of defined chirality which are separated in momentum space. Such a distinction is convenient to create a clear separation between intra- and intervalley scattering times and for maximizing the effect, but not necessary when discussing occupation changes local in momentum space. As pointed out in the introduction, the AGA appears because the dispersion is changed locally, for which the figure of merit is the (quantum) lifetime $\tau$ of the quasiparticle.
The corresponding kinetic equation is also known~\cite{Gao2019},
\begin{align}\label{eq:boltzmann2}
	\partial_t f_{\tilde{\varepsilon}}
	+(\bm{\tilde v}+e\bm{E})\partial_{\bm{r}}f_{\tilde{\varepsilon}}
	+e\bm{E}\partial_{\bm{p}}f_{\tilde{\varepsilon}}
	&=-\tfrac{f_{\tilde{\varepsilon}}-f^{(0)}}{\tau}
\end{align}
where the distribution function is evaluated at the shifted energy $\tilde\varepsilon=\varepsilon_0+e^2G_{ij}E_iE_j$, as is the velocity matrix element $\bm{\tilde v}$.
The current is defined as
\begin{align}\label{eq:currentdef}
	\bm{j}&=\int_{\bm{p}}
	f_{\tilde{\varepsilon}}
	\partial_{\bm{p}}\tilde\varepsilon.
\end{align}
The solution for $f$ reads 
\begin{align}
	f_{\tilde{\varepsilon}}&=
	f^{(0)}_{\tilde{\varepsilon}}
	+\frac{\tau e}{\hbar}\bm{E}\cdot
	\partial_{\bm{p}}f_{\tilde{\varepsilon}}
	+\frac{3\tau^2 e^2}{\hbar^2} E_iE_j
	\partial_{p_i}\partial_{p_j}f_{\tilde{\varepsilon}}.
\end{align}
Inserting this solution into the continuity equation yields $\bm{\nabla}\cdot \bm{j}=0$ and 
\begin{align}
\partial_t n=\int_{\bm{p}}\partial_t f_{\tilde{\varepsilon}}
=\int_{\bm{p}}\frac{\partial  f_{\tilde{\varepsilon}}}{\partial\tilde{\varepsilon}} \frac{ \partial\tilde{\varepsilon}}{\partial t}
&\approx 
-e^2\int_{\bm{p}}
\frac{\partial  
f_{\tilde{\varepsilon}}}{\partial\tilde{\varepsilon}}
\frac{G_{ij}E_iE_j}{\tau}
\label{eq:conti1}
\\
\delta n=\int_{\bm{p}}(f_{\tilde{\varepsilon}}-f^{(0)})
&\approx 
e^2\int_{\bm{p}}
\frac{\partial  f_{\tilde{\varepsilon}}}{\partial\tilde{\varepsilon}}
G_{ij}E_iE_j,
\label{eq:conti2}
\end{align}
In the first line, we made again use of the relaxation time approximation, while for $\delta n$ it was assumed that the change in dispersion is small compared to the sharpness of the Fermi surface, so that we can approximate $f^{(0)}_{\tilde{\varepsilon}}\approx 
f^{(0)}+\partial_{\varepsilon}f^{(0)}(\tilde{\varepsilon}-\varepsilon)$.
Eqs.~(\ref{eq:conti1},\ref{eq:conti2}) show that the steady-state solution for $f_{\tilde{\varepsilon}}$ fulfills the continuity equation but does not render it trivially zero.
Instead, charge is not conserved at order $\tau^{-1}$, and escapes via the collision term, analogously to case of a chiral anomaly. 
Compared to the latter, the anomaly is produced by the DDS  (Eq.~\eqref{eq:defDDS}). Like the Berry curvature, the DDS is a Fermi surface quantity which contains information about the entire bandstructure.
In conclusion, the continuity equation is not identically zero, and thus contains an anomaly.

As a necessary consistency check, we evaluated the entropy production at order $\tau^0$ to calculate the anomaly-related current. Joule heating is proportional to $(f-f^{(0)})^2$, so it is a nontrivial check if this recovers the same expression that is found by evaluating Eq.~\eqref{eq:currentdef}, which is linear in $f$. To this end, following Ref.~\cite{Son2013}, we estimate
\begin{align}
	\bm{j}\cdot\bm{E}&=T \dot S=
	T\int_{\bm{p}} \frac{(f_{\tilde{\varepsilon}}-f^{(0)})^2}{\tau f_{\tilde{\varepsilon}}(1-f_{\tilde{\varepsilon}})}
	\notag\\&
	=
	T\int_{\bm{p}} \frac{2(-\frac{\partial  
	f_{\tilde{\varepsilon}}}{\partial\tilde{\varepsilon}}G_{ij}E_iE_j)
	(\frac{\tau e}{\hbar}\bm{E}\cdot
	\partial_{\bm{p}}f_{\tilde{\varepsilon}})}{ \tau f_{\tilde{\varepsilon}}(1-f_{\tilde{\varepsilon}})}
	\notag\\&
	=
	\Bigl(-\frac{e^3}{\hbar}\int_{\bm{p}} 2G_{ij}E_iE_j
	\partial_{\bm{p}}f\Bigr)\cdot\bm{E}
\end{align}
where we kept in the numerator only terms of order $\tau^1$ and used that $\partial_{\varepsilon} f/(f(1-f))\equiv1/T$. The result is identical to Eq.~\eqref{eq:tau0}. This means firstly that the relaxation time approximation can be applied consistently throughout, the anomaly in the continuity equation is therefore not an artifact. Secondly, we find that there is no anomaly in the continuity equation for energy-momentum, at least not up to second order. 
This is commensurate with the covariant expression for the energy-momentum tensor, Eq.~\eqref{eq:covT}, which for the homogeneous bulk contains an anomaly only at order $\tau^{-2}$, and not at order $\tau^0$.

\begin{figure*}
    \centering
    \includegraphics[width=0.75\textwidth]{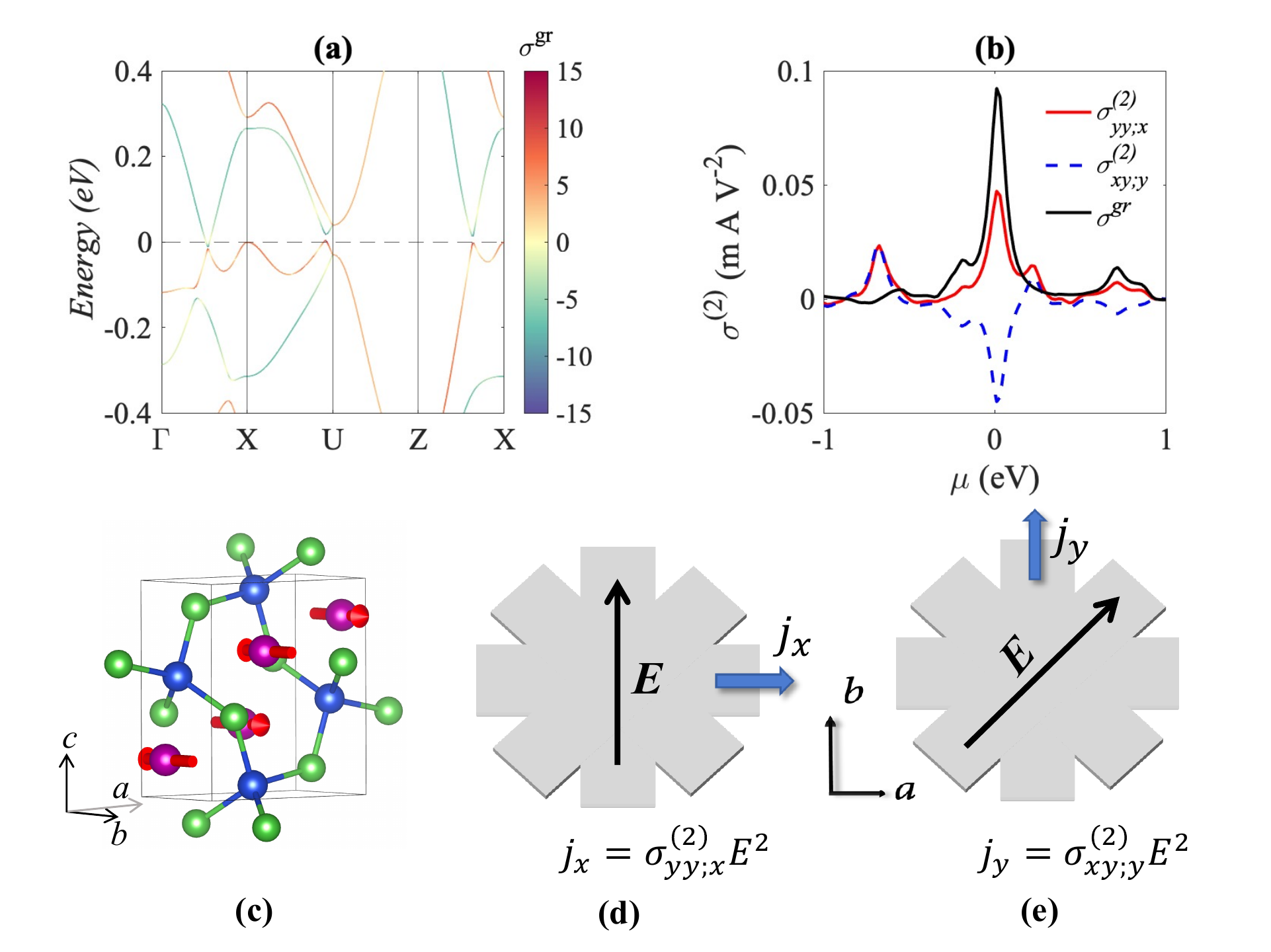}
    \caption{
    \textbf{Material proposal on the orthorhombic CuMnAs to detect the anomaly.}
    (a) Band structure of CuMnAs in the antiferromagnetic phase with massive Dirac points near the Fermi energy (zero). The color represents the band-resolved nonlinear conductivity due to gravitational anomaly, where the massive Dirac bands contribute significantly. 
    (b) The calculated nonlinear conductivities at different chemical potential ($\mu$). The gravitational anomaly contribution $\sigma^{gr}$ extracted from $\sigma^{(2)}_{yy;x}-\sigma^{(2)}_{xy;y}$. (c) The antiferromagnetic phase of CuMnAs where magnetic moments are along the $b$ axis.    (d)-(e) Two experimental setups to measure $\sigma^{(2)}_{yy;x}$ and $\sigma^{(2)}_{xy;y}$ separately. Here, $x,y$ align with the $a,b$ crystalline axes, respectively. In (e), the applied electric field (or bias voltage) orients 45$^\circ$ with respect to the $x$ direction. Additionally, the anomaly effect can also be extracted from the scaling of $\sigma^{(2)}$ in Eq.~\eqref{eq:scaling}. }
    \label{fig:fig3}
\end{figure*}
As mentioned, in a PT-symmetric material, $\sigma^{gr}\neq 0$. However, this cannot be due to the chiral anomaly, since the system under consideration is homogeneous and there is no magnetic field applied.
To understand the origin of the anomaly, we therefore investigate the general anomaly equations of chiral fermions in relativistic notation and for a curved spacetime, written in terms of the electromagnetic field strength $F_{\mu\nu}$, the Riemann curvature tensor $R^\mu_{\;\;\nu\lambda\kappa}$ and the Cristoffel symbol $\Gamma^{\mu}_{\;\;\nu\lambda}$. Here, greek indices denote covariant or contravariant spacetime indices, and we use the Einstein summation convention.
In field theory, the continuity (anomaly-) equations are considered in the so-called covariant formulation, which are then compared to the consistent formulation~\footnote{The word 'covariant' in this context has nothing to do with covariant coordinate indices.}. 
The covariant anomaly equations for the chiral current $J^{\mu}$ and the energy-momentum tensor $T^{\mu\nu}$ are well known~\cite{Jensen2013},
\begin{align}
    D_\mu J_{cov}^\mu&=
    \tfrac{3c_A}{4}\epsilon^{\kappa\sigma\alpha\beta}
    F_{\kappa\sigma}F_{\alpha\beta}+
    \tfrac{c_g}{4} \epsilon^{\kappa\sigma\alpha\beta}
    R^\nu_{\;\,\lambda\kappa\sigma}R^\lambda_{\;\,\nu\alpha\beta}
    \label{eq:covJ}
    \\
    D_\nu T_{cov}^{\mu\nu}&=
    F^\mu_{\;\;\nu} J_{cov}^\nu+\tfrac{c_g}{2}D_\nu(\epsilon^{\kappa\sigma\alpha\beta}
    F_{\kappa\sigma}R^{\mu\nu}_{\;\;\;\,\alpha\beta}),
    \label{eq:covT}
\end{align}
where $D_\mu$ denotes the covariant derivative, $c_A$ and $c_g$ are the anomaly coefficients of chiral anomaly and AGA, respectively.
The covariant current $J_{cov}$ is $U(1)$-gauge invariant and encodes the Fermi surface contributions to the current. However, this current does not necessarily preserve total charge, even after summing over both chiralities. This non-conservation is found for example in the presence of electromagnetic pseudofields~\cite{Zhou2013,Pikulin2016,Grushin2016,Ilan2020}. If this happens, total charge conservation is only restored after the addition of extra currents at the cutoff scale (band edge)~\cite{Gorbar2017,Gorbar2017b,Behrends2019a}. 
The addition of extra currents to $J_{cov}$ results in the consistent current $J_{cons}$, which is not $U(1)$-gauge invariant, but as mentioned does not lose any charge. The latter obeys modified anomaly equations
\begin{align}
    D_\mu J_{cons}^\mu&=
    \tfrac{c_A}{4}\epsilon^{\kappa\sigma\alpha\beta}
    F_{\kappa\sigma}F_{\alpha\beta}+
    (1-\alpha)\tfrac{c_g}{4} \epsilon^{\kappa\sigma\alpha\beta}
    R^\nu_{\;\,\lambda\kappa\sigma}R^\lambda_{\;\,\nu\alpha\beta}
    \label{eq:consJ}
    \\
    D_\nu T_{cons}^{\mu\nu}&=
    F^\mu_{\;\;\nu} J_{cons}^\nu
    -D_\lambda J_{cons}^\lambda A^\mu
    \notag\\&\quad
    -\alpha\tfrac{c_g g^{\mu\nu}}{2\sqrt{-g}}
    D_\lambda(\sqrt{-g}\epsilon^{\kappa\sigma\alpha\beta}
    F_{\kappa\sigma}\partial_\alpha\Gamma^{\lambda}_{\;\,\nu\beta}),
    \label{eq:consT}
\end{align}
where $\alpha$ parametrizes a Chern-Simons current $j_{CS}^{\mu}=\epsilon^{\mu\nu\kappa\sigma}[\Gamma^{\lambda}_{\;\rho\nu} D_\kappa\Gamma^\rho_{\;\lambda\sigma}+\tfrac{2}{3}\Gamma^{\lambda}_{\;\alpha\nu}\Gamma^\alpha_{\;\rho\kappa}\Gamma^\rho_{\;\lambda\sigma}]$ which is added to the action in the form of a contact term $-\alpha \int c_g A\wedge j_{CS}$. 

Let us analyze Eqs.~(\ref{eq:covJ}-\ref{eq:consT}) for a PT-symmetric system.
For the homogeneous bulk and without applied magnetic field it is $\bm{B}=\bm{B}_5=0$ and the chiral anomaly $\epsilon^{\kappa\sigma\alpha\beta}    F_{\kappa\sigma}F_{\alpha\beta}$ vanishes identically~\footnote{Note that the restriction to a homogenous current density is important. For inhomogeneous flow the chiral anomaly is present even in the absence of a magnetic field~\cite{Landsteiner2016}.}. 
Therefore, the anomaly in $\sigma^{gr}$ must necessarily be a mixed axial-gravitational anomaly.
Now, the presence of Fermi sea contributions as given by $j^\mu_{CS}$ depends on the microscopic properties of the system. If they were absent, it is $\alpha=0$ and thus $J_{cov}=J_{cons}$. Since $J_{cons}$ conserves total charge by construction, and $J_{cov}$ is independent of $\alpha$, the covariant current as induced by the AGA can conserve total charge without Chern-Simons currents. This has to be contrasted with the current if only the chiral anomaly terms are nonzero. In the latter case, invariably it is  $J_{cov}\neq J_{cons}$, which means that generically, additional currents must be imposed to restore total charge conservation.

Anomalies are related with a spectral flow in the band structure (i.e. they lead to a redistribution of occupation, cf.~\cite{Landsteiner2016}). The spectral flow for the AGA is implemented by deforming the band structure locally, which results in a (local) redistribution of carriers. 
Importantly, while the total number of carriers is conserved, the number of carriers at the Fermi level for a given chirality is not. In other words, between chiralities, the spectral flow transfers states from the Fermi sea to the Fermi surface, or vice versa. 
Finally, we comment on how the semiclassics connects to the general relativistic viewpoint. Because the equation of motion are normally not written for a curved background geometry, within the canonical semiclassical formulation the AGA instead enters through the time derivative of the density, which is consistent with its origin from an emergent curved space\emph{time}.

\emph{Experimental proposal.---} Of the nonlinear conductivities $\sigma^{(2)} = \sigma^{dr} + \sigma^{gr} + \sigma^{bc}$, $\sigma^{dr}$ and $\sigma^{gr}$ are nonzero only if the band structure breaks both spatial inversion symmetry ($\mathcal{P}$) and time-reversal ($\mathcal{T}$).  Conversely, if the combined symmetry $\mathcal{PT}$ still exists, the Berry curvature contribution $\sigma^{bc}$ vanishes~\cite{Zhang2019}.
In materials preserving $\mathcal{PT}$, also the intrinsic linear anomalous Hall effect vanishes and even extrinsic Hall effects due to skew scattering and side jump~\cite{Nandy2019,Xiao2019,Du2019} are suppressed~\cite{Watanabe2020}. Therefore, $\mathcal{PT}$ materials are ideal candidates to investigate the AGA through their intrinsic nonlinear conductivity.
On top of that, the only two remaining Hall contributions, $\sigma^{dr}$ and $\sigma^{gr}$, behave differently under permutation of the spatial indices, so that the Drude contribution $\sigma^{dr}$ can be subtracted away, for example, by taking a combination $\sigma^{(2)}_{yy;x}-\sigma^{(2)}_{xy;y}$.
Therefore, our proposal for an experimental detection of the AGA is depicted in Fig.~\refsub{fig:fig3}{d-e}.
It involves a subsequent measurement of, respectively, $\sigma^{(2)}_{yy;x}$ and $\sigma^{(2)}_{xy;y}$ in the same sample. 

We have calculated $\sigma^{(2)}$ and $\sigma^{gr}$ for the orthorohmbic phase of CuMnAs, which is antiferromagnetic and breaks $\mathcal{P}$ and $\mathcal{T}$ but preserves $\mathcal{PT}$~\cite{Emmanouilidou2017,Zhang2017}.
As shown in Fig.~\refsub{fig:fig3}{a}, this material features massive Dirac bands near the Fermi energy ($\mu=0$)~\cite{Tang2016,Emmanouilidou2017}. 
The second order conductivities $\sigma^{(2)}_{yy;x}$ and $\sigma^{(2)}_{xy;y}$ are shown in Fig.~\refsub{fig:fig3}{c}. Although they depend on band structure details, their difference $\sigma^{(2)}_{yy;x}-\sigma^{(2)}_{xy;y}$, the pure AGA contribution, exhibits a peak near massive Dirac points (Fig.~\refsub{fig:fig3}{c}).
Although it might be challenging to tune the chemical potential in a 3D material, it is enough to positively conclude that the anomaly is present by
detecting a finite difference $\sigma^{gr}=\sigma^{(2)}_{yy;x}-\sigma^{(2)}_{xy;y}$. 

As discussed above, the anomaly term $\sigma^{gr}$ is independent of $\tau$, similar to the intrinsic anomalous Hall effect. Thus, we expect $\sigma^{gr}$ is insensitive to temperature variation or the longitudinal conductivity, even though $\sigma^{(2)}_{yy;x}$ and $\sigma^{(2)}_{xy;y}$ are $\tau$ dependent. By switching the $x,y$ axes, one can also probe another anomaly term 
$\sigma^{gr}=\sigma^{(2)}_{xx;y}-\sigma^{(2)}_{yx;x}$, which exhibits a similar peak around gapped Dirac points. Similar measurements can also be performed in the $yz$ and $xz$ planes to probe AGA, which involves differences between cyclically permuted $\sigma^{(2)}$ (see Appendix). Here, the $xy$ and $xz$ planes exhibits larger AGA signals than the $yz$ plane.
In addition, we note that $\sigma^{(2)}$ is sensitive to magnetic order and crystal structure of the material. Both $\sigma^{(2)}$ and $\sigma^{gr}$ will vanish if the material recovers inversion symmetry above the Neel temperature. 

The scaling of $\sigma^{(2)}$ to the linear conductivity $\sigma$ is different from the Berry-curvature induced nonlinear anomalous Hall effect~\cite{Kang2019,Ma2019}. 
For the $\mathcal{PT}$ system, $\sigma^{(2)}$ has two terms which scale differently with the relaxation time $\tau$, in the form of $\sigma^{dr} \propto \tau^2$ and $\sigma^{gr} \propto \tau^{0}$.
Since $\sigma \propto \tau$, we obtain the following relation,
\begin{align}
    \frac{E_{\perp}}{E_{||}^2} = \frac{\sigma^{(2)}}{\sigma} =
    \eta_1 \sigma + \frac{\eta_{2}}{\sigma},
    \label{eq:scaling}
\end{align}
where $E_{\perp}$ and $E_{||}$ are transverse and longitudinal electric field, respectively. Here, $\eta_1 = \frac{\sigma^{dr}}{\sigma^2}$ and $\eta_2 = \sigma^{gr}$. We emphasize that $\eta_2$  directly probes the AGA effect, even without subtracting the permuted $\sigma^{(2)}$ partner. 

To estimate the strength of the nonlinear Hall effect, we adopt the peak value of $\sigma^{(2)}_{yy;x}\approx 0.05 ~\mathrm{mA V^{-2}}$ in Fig.~\refsub{fig:fig3}{b}, which is in the same order of magnitude as $\sigma^{gr}$. 
Assuming a reasonable electric field strength $E_y = 1~\mathrm{V\,cm^{-1}}$, the effective anomalous Hall conductivity is $\sigma^{AHE} = \sigma^{(2)}_{yy;x} E_y \approx 50 ~ \mathrm{\Omega^{-1} \mu m^{-1}} $, which is comparable to that of some Fe thin-films~\cite{Nagaosa2010}. The effective Hall angle is then $\gamma = \frac{\sigma^{AHE} }{ \sigma} \approx 0.5\% $, where we used $\sigma \approx 10^4 ~\mathrm{S\,cm^{-1}}$ according to Ref.~\cite{Zhang2017}.

\emph{Concluding remarks.---}
We have shown that in systems without time reversal and spatial inversion, an anomalous electrical current appears at second order in the electric field due to the axial-gravitational anomaly. 
We suggest a straightforward multi-contact geometry to measure this current in an all-electrical setup. 

We note that for a system with a number of Weyl cones at the Fermi energy, the phenomenology presented here can be translated into a time-dependent effect of a pseudoeletric field, whereby the latter is a vector quantity which constitutes a simplified parametrization of the effect of the curvature tensor that holds due to the linear dispersion near the nodal points. This formalism will be presented elsewhere.

The observations put forth here have quite far-reaching implications for the theory of quantum transport. It allows, for the first time, to probe in the electrical conductivity the nontrivial emergent spacetime in which the electrons move when traversing a periodic lattice. 
This emergent curved spacetime is consistent with a classical picture of transport of deformable wavepackets flexing and bending while they squeeze through the lattice atoms~\cite{Holder2021}. 
These results offer the exciting perspective that dynamical effects of motion in a Riemannian curved spacetime could be engineered and accessed in a genuine bulk setting, and not by strain engineering~\cite{Zhou2013,Zubkov2015,Weststroem2017,Cortijo2016,Volovik2016,Guan2017,Zubkov2018}.
It also opens up a new route to investigate the interplay of anomalies, Bardeen-Zumino currents and their role in quantum field theories in curved spacetime in a lattice, similar to recent developments regarding the chiral anomaly~\cite{Ilan2020}.
More concretely, a generalization of the program undertaken here, to compare perturbation theory and semiclassics should yield an explicit expression for the effective Riemann curvature tensor in the emergent spacetime in a lattice and might allow a derivation of the Luttinger approach for thermal transport~\cite{Luttinger1964}.

\begin{acknowledgments}
We thank 
Erez Berg,
Tabea Heckenthaler,
Johannes S. Hofmann,
Karl Landsteiner and
Raquel Queiroz
for fruitful discussions.
B.Y.\ acknowledges the financial support by the European Research Council (ERC Consolidator Grant No. 815869, ``NonlinearTopo'') and Israel Science Foundation (ISF No. 2932/21). 
R.I.\ was supported by the Israel Science Foundation (ISF No. 1790/18).
\end{acknowledgments}

%

\newpage
\onecolumngrid
\appendix
\begin{center}
\large
\textbf{Appendix}
\end{center}
 \setcounter{equation}{0}
 \renewcommand{\theequation}{A\arabic{equation}}
 \setcounter{figure}{0}
 \renewcommand{\thefigure}{A\arabic{figure}}
\section{Probing the anomaly in different crystallographic planes}
As stated in the main text, other combinations of the principal axes $x,y,z$ aligned with the $a,b,c$ directions permit measurement of $\sigma^{\textrm{gr}}$. For completeness, we provide additional calculations, the presence of the anomaly for other spatial indices. 
\begin{figure*}[h]
    \centering
    \includegraphics[width=0.8\textwidth]{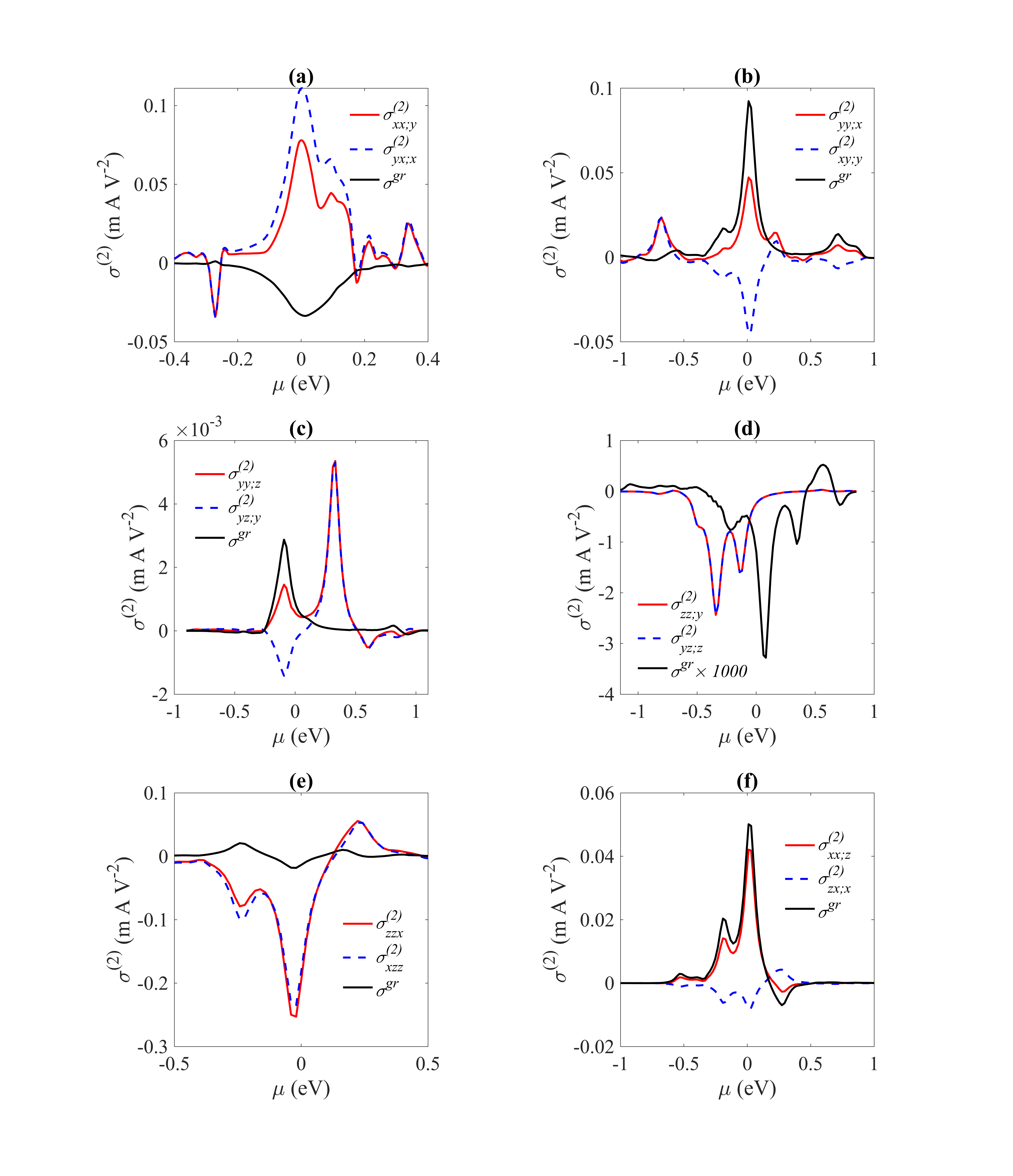}
    \caption{ Nonlinear conductivity $\sigma^{ab;c}$ producing non-zero $\sigma^{\textrm{gr}}$ in the $xy, yz, xz$ planes. The anomaly is generated by the subtraction of $\sigma^{gr} = \sigma_{aa;b}^{(2)} - \sigma_{ba;a}^{(2)}$. In (d), for legibility, $\sigma^{gr}$ is scaled here by $1000$.}
    \label{fig:fig4}
\end{figure*}
\\
In the event the principle axes are misaligned with respect to the crystallographic directions $a,b,c$, an interpolation is possible through the projection of the directions $\hat{x}, \hat{y}, \hat{z}$. We may define $\cos(\gamma_i) = \hat{r}_i \cdot a_i$, where $a_i = \lbrace a, b,c\rbrace$ as the angle between the principle axes and the crystallographic directions. Next, it is possible to decompose a current in an arbitrary direction, $j^{x} = \sum_{i,j} A_{ia} A_{jb}A_{xc} \sigma_{ab;c}^{\textrm{(2)}}E_a E_b$. Here $A_{ia}$ is the rotation matrix of the crystallographic direction $a$ about the $i$th principle axis, by the angle $\gamma_i$ defined above. We stress that the choice of coordinate system will not affect the conclusion regarding the presence of an anomaly in the system; a misaligned coordinate system could only \textit{admix} components from different crystallographic directions. When heating the sample above its Neel temperature, we expect the anomalous signal to vanish completely, thus nulling the response in an arbitrary direction as well. 

\section{Computational Methods}
Ab initio calculations were performed on orthorhomboic CuMnAs using the full-potential local-orbital minimum-basis code \cite{Koepernik1999}. A $12\times 12 \times 12$ reciprocal lattice grid was used to obtain the ground state wavefunctions and energies, which were then projected on 144 atomic-site projected Wannier functions. Ground state properties were converged with a tolerance of $10^{-6} \textrm{eV}$ for the total energy. All momentum space integrals were carried out on a $350 \times 350 \times 350$ grid in the first Brillouin zone, where convergence was verified when incremental increases in the grid size amounted to less than $5\%$ difference in the integrated value.
The anti-ferromagnetic spin configuration of the $Mn$ atoms was implemented in accordance with experimental findings on the commensurate AFM (c-AFM) structure in orthorhombic CuMnAs \cite{Emmanouilidou2017, Zhang2017}, with spins oriented along the $b$-axis. We recover the experimental result that the c-AFM order breaks the screw symmetry $S_{2z}$, while also breaking the $R_y$ nonsymmorphic reflection symmetry. The symmetry breaking is manifest in the appearance of massive Dirac cones in Fig.~\ref{fig:fig3}(a) along the $\Gamma-X, X-U, Z-X$ lines, unlike the case when spins are aligned with the $c$ direction, and $R_y$ and $S_{2z}$ protect the crossing points \cite{Tang2016}. Our observed gap of $\Delta = 5 \textrm{meV}$ along $\Gamma-X$ is in agreement with previously reported values \cite{Emmanouilidou2017}. 
The non-magnetic structure of orthorhombic CuMnAs is that of space group 62 (Pnma), which includes the inversion. Upon the introduction of c-AFM order, the space group is reduced to P$2_1$m (11), which breaks the original $S_{2z}$ screw symmetry and $R_y$, as noted above. Besides the conductivities shown in Fig.~\ref{fig:fig4}, two other components are symmetry-permitted which do not contribute to the anomaly: $\sigma^{(2)}_{xx;x}$ and $\sigma^{(2)}_{zz;z}$. We note that the longitudinal nonlinear terms can lead to nonreciprocal, unidrectional magnetoresistance in transport due to breaking of both inversion and time-reversal symmetries \cite{Liu2020b}. Signs of $\sigma^{(2)}_{xx;x}$ and $\sigma^{(2)}_{zz;z}$ can probe the orientation or Neel Vector of the antiferromagnetic order. 

\end{document}